\documentclass[letterpaper, 10 pt, conference]{ieeeconf}  

\IEEEoverridecommandlockouts                              

\usepackage{lipsum} 
\usepackage{scrextend} 
\usepackage[pdftex]{graphicx} 
\usepackage{siunitx} 
\usepackage{multirow} 
\usepackage{array} 
\let\labelindent\relax\usepackage{enumitem} 
\usepackage{color} 
\usepackage{colortbl} 
\usepackage{pifont} 
\usepackage{amsmath} 
\usepackage{amssymb}
\usepackage{dblfloatfix} 
\usepackage[ruled,vlined,resetcount]{algorithm2e}
\usepackage[caption=false,font=footnotesize]{subfig}
\usepackage{tikz}
\usetikzlibrary{shapes, arrows, fit, calc, positioning, automata, decorations.markings, backgrounds}
\tikzset{>=latex}

\makeatletter
\let\NAT@parse\undefined
\makeatother
\PassOptionsToPackage{hyphens}{url}
\usepackage[colorlinks=true,linkcolor=blue,urlcolor=blue,citecolor=blue,anchorcolor=blue]{hyperref} 

\graphicspath{{figures/}} 

\newtheorem{remark}{Remark}
\newtheorem{assumption}{Assumption}

\title{\LARGE \bf
Dynamic System Emulation: Fixed Wing Dynamics on a Multicopter
}

\author{Abdelhakim Amer$^{1}$ and Andriy Sarabakha$^{1}$
\thanks{This research was supported by the Aarhus University Research Foundation.}
\thanks{$^{1}$Abdelhakim Amer and Andriy Sarabakha are with the Department of Electrical and Computer Engineering, Aarhus University, Denmark. {\tt\small \{abdelhakim, andriy\}@ece.au.dk}}%
}

\begin{document}

\bstctlcite{IEEEexample:BSTcontrol}

\maketitle
\thispagestyle{empty}
\pagestyle{empty}

\begin{abstract}

This work presents a control framework that enables a multicopter equipped with a two-axis gimbal to emulate the flight dynamics of a fixed-wing aircraft. The goal is to provide an operationally simple platform for training and simulation that avoids the aerodynamic constraints of fixed-wing vehicles, such as minimum airspeed and nonholonomic constraints.
A state-input mapping between the two platforms is derived using dynamic feedback linearization. 
The framework is evaluated on representative fixed-wing manoeuvres. 
Results show high-fidelity emulation under nominal conditions. While evaluations are conducted in simulation, the approach establishes a practical path toward hardware deployment for pilot training, autonomy research, and controller benchmarking.

\end{abstract}


\section{Introduction}

Mimicking the behaviour of one dynamic system with another, broadly termed emulation, is a powerful tool in engineering and research. These approaches can be classified into a taxonomy based on their connection to the real world. The first category is virtual simulation, in which a computer model replicates a system's dynamics and sensory environment~\cite{shah2017airsim, amer2023unav}. While valuable for training algorithms, these systems suffer from the well-known ``sim-to-real'' gap, as models trained on simulated data often fail when deployed in the real world~\cite{gao2020realitygap, aljalbout2025reality}. 
The second category is pilot-in-the-loop~(PIL) simulation, which combines physical motion platforms with virtual environments, primarily for human operator training. Such systems~\cite{stroosma2003using, edvani1997simona}, provide realistic motion cues to pilots while rendering either computer-generated visuals or pre-recorded imagery. However, PIL systems remain fundamentally limited for algorithm validation: they cannot provide real-time sensor data that responds dynamically to the system's current state and actions.

This work proposes a third paradigm: physical-to-physical system emulation. In this approach, one highly manoeuvrable, low-cost physical system, i.e a multicopter equipped with a camera on a gimbal,  is controlled to dynamically replicate both the motion and camera viewpoint of another, more constrained system, a fixed-wing aircraft. 
Fixed-wing aircraft are operationally demanding due to aerodynamic constraints such as minimum airspeed requirements and stall risks. In contrast, a multicopter offers affordability and superior maneuverability (including hover and vertical takeoff), making it an ideal emulation platform. This approach is particularly valuable for validating algorithms at lower technology readiness levels~\cite{smith2022emulation} and enables flexible, repeatable data collection for perception tasks by providing real-time sensory data in real-world environments.

However, achieving this emulation, sometimes is not straightforward. The dynamics of fixed-wing aircraft and multicopters differ fundamentally: fixed-wing systems exhibit nonholonomic constraints, aerodynamic stall behaviors, and actuation models that have no direct analogue in multicopter dynamics~\cite{yuan2025design, wu2019modeling}. This raises two critical questions: Is mimicry theoretically possible given these fundamental differences? And if so, what control techniques can realize it in practice?
Addressing the first question requires formal analysis of feasibility. Recent work has introduced similarity frameworks that define homeomorphic maps to transform system trajectories, quantifying emulation fidelity via cost functions~\cite{wang2023similarity}. This concept of similarity degree provides a formal method for determining whether mimicry is theoretically achievable and how well it can be approximated. Once mimicry feasibility is established, the second question remains: mapping these disparate behaviors from the fixed-wing (primary) system to the multicopter (secondary) system requires sophisticated control techniques.
While model predictive control (MPC) is a popular approach for handling constraints~\cite{amer2023visual, amer2025}, its high computational cost can be prohibitive for high degrees of freedom, real-time applications on embedded hardware. An alternative and highly effective strategy is dynamic feedback linearization~(DFL). DFL transforms complex nonlinear system dynamics into a simpler, linear form, often by dynamically extending the state space. This allows for systematic and computationally efficient trajectory tracking, making it a powerful tool for managing systems with complex relative degrees, as is the case in the emulation problem~\cite{pshuniak2018relative}.

The objective of this work is to bridge the gap between theoretical similarity and practical implementation by designing and validating a DFL-based framework that enables a multicopter with a two-axis gimbal to emulate the dynamics and sensor viewpoint of a fixed-wing aircraft. Specifically, the contributions of this work are: (1) a proof of the multicopter-gimbal's camera-pose controllability and mimicry capability; (2) the design of DFL-based control laws to mimic fixed-wing camera pose trajectories; and (3) validation of the complete framework through simulations of increasingly complex manoeuvres.

This manuscript is organized as follows. Section~\ref{sec:preliminaries} summarizes the preliminary information about the aircraft models. Section~\ref{sec:proof} provides the controllability and mimicry proofs. Section~\ref{sec:method} details the proposed DFL-based imitation method. Section~\ref{sec:results} provides validation results, and Section~\ref{sec:conclusions} concludes the work.

\section{Preliminaries}
\label{sec:preliminaries}

Quaternions $\mathbf{q} \in \mathbb{H}_1$ are scalar-first, their conjugate are $\bar{\mathbf{q}} \in \mathbb{H}_1$, and $\mathbf{R}(\mathbf q) \in SO(3)$ denotes the corresponding rotation matrix. Also, let $\mathcal{F}_\text{W}$ be the inertial world frame. 

\subsection{Fixed-Wing Aircraft Dynamics}

\begin{figure}[!b]
\centering
\includegraphics[width=0.9\columnwidth, trim={0 0 1cm 0}, clip]{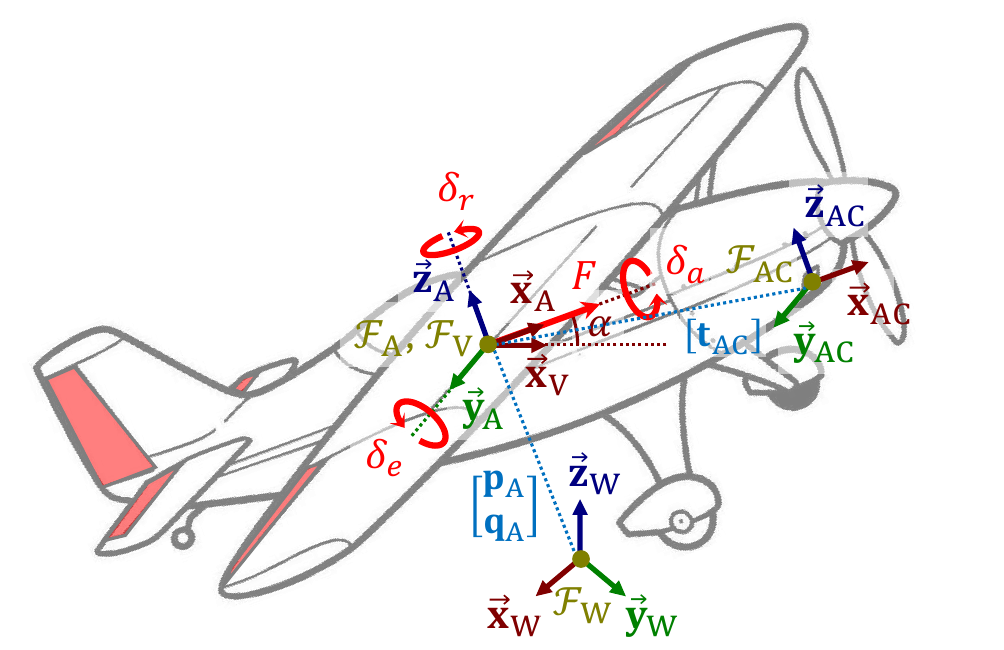}%
\caption{Fixed-wing aircraft shown with its reference frames, control inputs (in red) and output variables (in blue).}
\label{fig:plane}
\end{figure}

Fixed-wing aircraft (in Fig.~\ref{fig:plane}) are underactuated dynamic systems. 
Let $\mathcal{F}_\text{A}$ be the aircraft's body frame. The aircraft's pose is given by its position $\mathbf p_\text{A} = \begin{bmatrix} x_\text{A} & y_\text{A} & z_\text{A} \end{bmatrix}^\top \in \mathbb{R}^3$ in $\mathcal F_\text{W}$ and a quaternion $\mathbf q_\text{A} \in \mathbb{H}_1$ mapping $\mathcal F_\text{A} \to \mathcal F_\text{W}$ with $\phi_\text{A}$, $\theta_\text{A}$ and $\psi_\text{A}$ being roll, pitch and yaw angles, respectively. 
Aircraft's linear and angular velocities in $\mathcal{F}_\text{A}$ are $\mathbf{v}_\text{A} \in \mathbb{R}^3$ and $\boldsymbol{\omega}_\text{A} \in \mathbb{R}^3$, respectively.
The aircraft's state is
\begin{equation}
\mathbf{x}_\text{A} = \begin{bmatrix} \mathbf{p}_\text{A} & \mathbf{q}_\text{A} & \mathbf{v}_\text{A} & \boldsymbol{\omega}_\text{A} \end{bmatrix}^\top,
\label{eq:plane_x}
\end{equation}
while the output is
\begin{equation}
\mathbf{y}_\text{A} = \begin{bmatrix} \mathbf{p}_\text{A}^\top & \mathbf{q}_\text{A}^\top \end{bmatrix}.
\label{eq:plane_y}
\end{equation}

Commonly, the control input to a fixed-wing aircraft is 
\begin{equation}
\mathbf{u}_\text{A} = \begin{bmatrix} F & \delta_a & \delta_e & \delta_r \end{bmatrix}^\top,
\label{eq:plane_u}
\end{equation}
where $F$ is the thrust generated by the engines, $\delta_a$, $\delta_e$, and $\delta_r$ are  aileron, elevator, and
rudder deflections.
Aerodynamic forces acting on a fixed-wing aircraft defined in the wind frame $\mathcal{F}_\text{V}$ are $\mathbf{f}_\text{aero} = \begin{bmatrix} -D & Y & L \end{bmatrix}^\top$, with $D \propto v^2$ is drag, $Y$ is side force, $L \propto v^2$ is lift, and $v$ is airspeed. The total force acting on the fixed-wing aircraft in $\mathcal{F}_\text{A}$ is $\mathbf{f}_\text{A} = \mathbf{R}(\alpha) \mathbf{f}_\text{aero} + \begin{bmatrix} F & 0 & 0 \end{bmatrix}^\top + \mathbf{R}\left(\mathbf{q}_\text{A}\right)^\top \begin{bmatrix} 0 & 0 & -g \end{bmatrix}^\top$, where $\alpha$ is the angle of attack, and $g$ is the gravitational constant.
The total torque acting on the fixed-wing aircraft in $\mathcal{F}_\text{A}$ is $\boldsymbol{\tau}_\text{A} = \mathbf{B}_\tau \begin{bmatrix} \delta_a & \delta_e & \delta_r \end{bmatrix}^\top$, where $\mathbf{B}_\tau$ is the control-effectiveness map.
Using the Newton-Euler equations, the kinematics and dynamics of a fixed-wing aircraft are:
\begin{equation}
f_\text{A}:
\begin{cases}
\dot{\mathbf{p}}_\text{A} = \mathbf{R}\left(\mathbf{q}_\text{A}\right) \mathbf{v}_\text{A} \\
\dot{\mathbf{q}}_\text{A} = \frac{1}{2} \mathbf{G}\left(\mathbf{q}_\text{A}\right) \boldsymbol{\omega}_\text{A} \\
\dot{\mathbf{v}}_\text{A} = -\boldsymbol{\omega}_\text{A} \times \mathbf{v}_\text{A} + \frac{1}{m_\text{A}} \mathbf{f}_\text{A} \\
\dot{\boldsymbol{\omega}}_\text{A} = -\left(\mathbf{J}_\text{A}\right)^{-1} \boldsymbol{\omega}_\text{A} \times (\mathbf{J}_\text{A} \boldsymbol{\omega}_\text{A}) + \left(\mathbf{J}_\text{A}\right)^{-1} \boldsymbol{\tau}_\text{A}
\end{cases},
\label{eq:plane_dynamics}
\end{equation} 
in which 
$\mathbf{G}\left(\mathbf{q}\right) \in \mathbb{R}^{4 \times 3}$ 
is the quaternion kinematics matrix, $m_\text{A}$ is aircraft's mass, and $\mathbf{J}_\text{A} \in \mathbb{R}^{3 \times 3}$ is matrix of inertia.

\subsection{Multicopter with 2-Axis Gimbal Dynamics}

\begin{figure}[!b]
\centering
\includegraphics[width=0.9\columnwidth, trim={0 0 0 1cm}, clip]{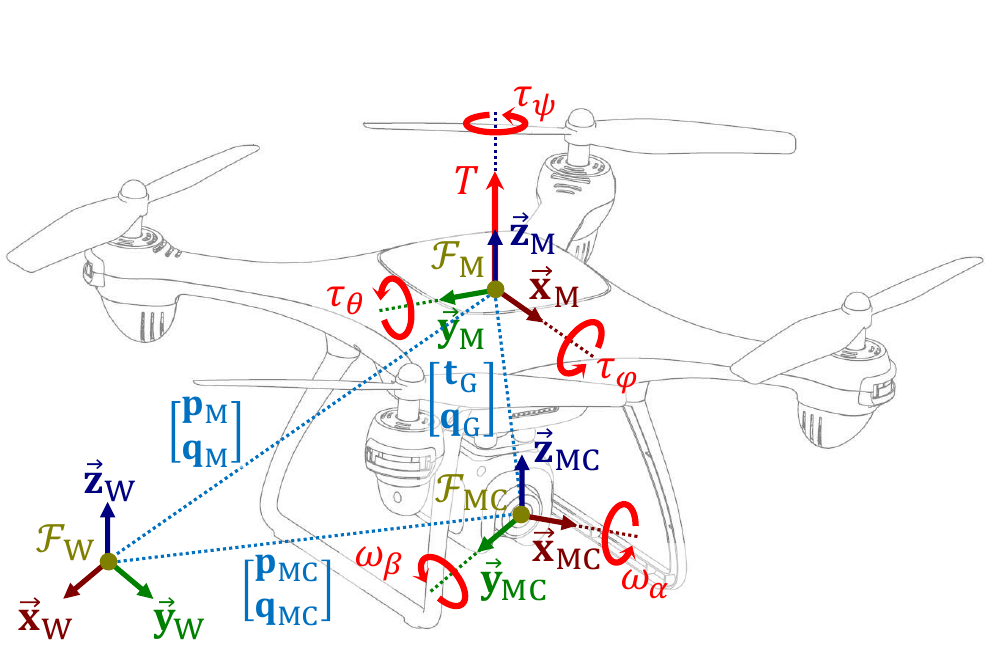}%
\caption{Multicopter with two-axis gimbal shown with its reference frames, control inputs (in red) and output variables (in blue).}
\label{fig:quadcopter}
\end{figure}

Multicopters (in Fig.~\ref{fig:quadcopter}) are also underactuated dynamic systems. 
Let $\mathcal{F}_\text{M}$ be the multicopter's body frame. The multicopter's pose is given by its position $\mathbf p_\text{M} = \begin{bmatrix} x_\text{M} & y_\text{M} & z_\text{M} \end{bmatrix}^\top \in \mathbb{R}^3$ in $\mathcal F_\text{W}$ and a quaternion $\mathbf q_\text{M} \in \mathbb{H}_1$ mapping $\mathcal F_\text{M} \to \mathcal F_\text{W}$ with $\phi_\text{M}$, $\theta_\text{M}$ and $\psi_\text{M}$ being roll, pitch and yaw angles, respectively. 
Multicopter's linear and angular velocities in $\mathcal{F}_\text{M}$ are $\mathbf{v}_\text{M} \in \mathbb{R}^3$ and $\boldsymbol{\omega}_\text{M} \in \mathbb{R}^3$, respectively.
The multicopter is equipped with a camera mounted on a 2-axis roll-pitch gimbal with joint angles $\phi_\text{G}$ and $\theta_\text{G}$ and quaternion $\mathbf{q}_\text{MC} = \mathbf{q}_x(\phi_\text{G}) \otimes \mathbf{q}_y(\theta_\text{G}) \in \mathbb{H}_1$, which maps from the multicopter's camera frame $\mathcal{F}_\text{MC}$ to $\mathcal{F}_\text{M}$. Let the camera lever arm in $\mathcal{F}_\text{M}$ be $\mathbf{t}_\text{G} \in \mathbb{R}^3$.
The state of a multicopter with a two-axis gimbal is 
\begin{equation}
\mathbf{x}_\text{M} = \begin{bmatrix} \mathbf{p}_\text{M} & \mathbf{q}_\text{M} & \mathbf{v}_\text{M} & \boldsymbol{\omega}_\text{M} & \mathbf{q}_\text{G} \end{bmatrix}^\top,
\label{eq:quad_x}
\end{equation}
while the output is the camera's pose:
\begin{equation}
\mathbf{y}_\text{MC} = \begin{bmatrix} \mathbf{p}_\text{M}^\top + \left( R\left(\mathbf{q}_\text{M}\right) \mathbf{t}_\text{G} \right)^\top & \left( \mathbf{q}_\text{M} \otimes \mathbf{q}_\text{G} \right)^\top \end{bmatrix}.
\label{eq:gimbal_y}
\end{equation}

Commonly, the control input to a multicopter is 
\begin{equation}
\mathbf{u}_\text{M} = \begin{bmatrix} T & \tau_\phi & \tau_\theta & \tau_\psi \end{bmatrix}^\top,
\label{eq:quad_u}
\end{equation}
where $T$ is total thrust, $\tau_\phi$, $\tau_\theta$, and $\tau_\psi$ are torques around $x$-, $y$- and $z$-axes, respectively. Control inputs to the gimbal are
\begin{equation}
\mathbf{u}_\text{G} = \begin{bmatrix} \dot\theta_\text{G} & \dot\psi_\text{G} \end{bmatrix}^\top.
\label{eq:gimbal_u}
\end{equation}
The total force acting on the multicopter in $\mathcal{F}_\text{M}$ is $\mathbf{f}_\text{M} = \begin{bmatrix} 0 & 0 & T \end{bmatrix}^\top + \mathbf{R}\left(\mathbf{q}_\text{M}\right)^\top \begin{bmatrix} 0 & 0 & -g \end{bmatrix}^\top$; while the total torque in $\mathcal{F}_\text{M}$ is $\boldsymbol{\tau}_\text{M} = \begin{bmatrix} \tau_\phi & \tau_\theta & \tau_\psi \end{bmatrix}^\top$.
So, the kinematics and dynamics of a multicopter with a gimbal are:
\begin{equation}
f_\text{M}:
\begin{cases}
\dot{\mathbf{p}}_\text{M} = \mathbf{R}\left(\mathbf{q}_\text{M}\right) \mathbf{v}_\text{M} \\
\dot{\mathbf{q}}_\text{M} = \frac{1}{2} \mathbf{G}\left(\mathbf{q}_\text{M}\right) \boldsymbol{\omega}_\text{M} \\
\dot{\mathbf{v}}_\text{M} = -\boldsymbol{\omega}_\text{M} \times \mathbf{v}_\text{M} + \frac{1}{m_\text{M}} \mathbf{f}_\text{M} \\
\dot{\boldsymbol{\omega}}_\text{M} = -\left(\mathbf{J}_\text{M}\right)^{-1} \boldsymbol{\omega}_\text{M} \times (\mathbf{J}_\text{M} \boldsymbol{\omega}_\text{M}) + \left(\mathbf{J}_\text{M}\right)^{-1} \boldsymbol{\tau}_\text{M} \\
\dot{\mathbf{q}}_\text{MC} = \frac{1}{2} \mathbf{G}\left(\mathbf{q}_\text{MC}\right) \mathbf{u}_\text{G}
\end{cases},
\label{eq:quad_dynamics}
\end{equation} 
in which $m_\text{M}$ is the mass of multicopter with gimbal, and $\mathbf{J}_\text{M} \in \mathbb{R}^{3 \times 3}$ is matrix of inertia.

\section{Camera-Pose Controllability and Mimicry on $SE(3)$}
\label{sec:proof}

Assume the fixed-wing carries a camera that is body-aligned and located at a fixed lever arm $\mathbf{t}_\text{AC} \in \mathbb{R}^3$ expressed in $\mathcal{F}_\text{A}$.
Then, the fixed-wing camera pose is
\begin{equation}
\mathbf{y}_\text{AC} = \begin{bmatrix} \mathbf{p}_\text{AC}^\top & \mathbf{q}_\text{AC}^\top \end{bmatrix}
= \begin{bmatrix} \big( \mathbf{p}_\text{A} + \mathbf{R}\left( \mathbf{q}_\text{A} \right) \mathbf{t}_\text{AC} \big)^\top & \mathbf{q}_\text{A}^\top \end{bmatrix}.
\label{eq:plane_camera_pose}
\end{equation}



We expose (i) a translational ``virtual'' input $\,\mathbf{a}_\text{MC} \in \mathbb{R}^3$ realized by the standard $SE(3)$ inner loop (third line of~\eqref{eq:quad_dynamics}), and (ii) rotational inputs $\dot{\boldsymbol{\Theta}}_\text{MC} = \begin{bmatrix} \dot\phi_\text{G} & \dot\theta_\text{G} & \dot\psi_\text{M} \end{bmatrix}^\top$ consisting of gimbal roll and pitch rates and the multicopter’s free body yaw rate around $\vec{\mathbf{z}}_\text{M}$, which does not alter the thrust direction.

From \eqref{eq:gimbal_y}, the position of the multicopter's camera in the $\mathcal{F}_\text{W}$ is 
\begin{equation}\mathbf{p}_\text{MC} = \mathbf{p}_\text{M} + \mathbf{R}(\mathbf{q}_\text{M}) \mathbf{t}_\text{G},\
\label{eq:p_mc}
\end{equation}
while the rotation is 
\begin{equation}
\mathbf{q}_\text{MC} = \mathbf{q}_\text{M} \otimes \mathbf{q}_\text{G}. 
\label{eq:q_mc}
\end{equation}
Then, by \eqref{eq:p_mc}
\begin{equation}
\ddot{\mathbf{p}}_\text{MC} = \mathbf{a}_\text{M} + \boldsymbol{\Delta}(\mathbf{p}_\text{MC}, \mathbf{q}_\text{M}, \boldsymbol{\omega}_\text{M}, \dot{\boldsymbol{\omega}}_\text{M}),
\label{eq:pc_ddot}
\end{equation}
where $\boldsymbol\Delta(\mathbf{p}_\text{MC}, \mathbf{q}_\text{M}, \boldsymbol{\omega}_\text{M}, \dot{\boldsymbol{\omega}}_\text{M})$ collects the known lever-arm coupling terms, and by \eqref{eq:q_mc}
\begin{equation}
\dot{\mathbf{q}}_\text{MC} = \frac{1}{2} \mathbf{G}(\mathbf{q}_\text{MC}) \boldsymbol{\omega}_\text{MC},
\label{eq:qc_dot}
\end{equation}
where $\boldsymbol{\omega}_\text{MC} = \mathbf{S}(\phi_\text{G}, \theta_\text{G}) \dot{\boldsymbol{\Theta}}_\text{MC}$ with the input-twist matrix
\begin{equation}
\mathbf{S}(\phi,\theta)=
\begin{bmatrix}
1 & 0 & -\sin\theta\cos\phi\\
0 & 1 & -\sin\phi\\
0 & 0 & \ \cos\theta\cos\phi
\end{bmatrix},
\label{eq:S_matrix}
\end{equation}
whose determinant is 
\begin{equation}
\det \mathbf{S}(\phi,\theta)=\cos\theta\,\cos\phi.
\label{eq:S_determinant}
\end{equation}

\begin{remark}
Singularities for a 2-axis gimbal occur at $\cos\phi_\text{G}\cos\theta_\text{G} = 0$.
\end{remark}

\begin{assumption}[Non-Singularity]
\label{ass:non_singularity}
The multicopter gimbal attitude stays away from the singular set: $\phi_\text{G} \neq \pm \frac{\pi}{2} \land \theta_\text{G} \neq \pm \frac{\pi}{2}$.
\end{assumption}

\begin{remark}
This assumption can be relaxed by using a planner which will avoid the singular set.
\end{remark}

\subsection{Output Full Actuation \& STLC on $SE(3)$}
\label{ssec:controllability}

Consider the block-triangular Jacobian from inputs to camera acceleration and twist:
\begin{equation}
\label{eq:J_block}
\begin{bmatrix} \ddot{\mathbf{p}}_\text{MC} \\ \boldsymbol{\omega}_\text{MC} \end{bmatrix} = 
\underbrace{\begin{bmatrix} \mathbf{I}_3 & \mathbf{0} \\ \mathbf{0} & \mathbf{S}(\phi_\text{G},\theta_\text{G})\end{bmatrix}}_{\mathbf{J}} \begin{bmatrix} \mathbf{a}_\text{MC} \\ \dot{\boldsymbol{\Theta}}_\text{MC} \end{bmatrix} + \begin{bmatrix} \boldsymbol\Delta \\ \mathbf{0} \end{bmatrix}.
\end{equation}

\paragraph{Translational rank}
Thrust vectoring yields $\mathbf{a}_\text{MC} = -g \mathbf{e}_3 + \frac{T}{m_\text{M}} \mathbf{b}_3$; variations $\delta \mathbf{a}_\text{MC} = \frac{\delta T}{m_\text{M}} \mathbf{b}_3 + \frac{T}{m_\text{M}} \delta \mathbf{b}_3$ span $\mathbb{R}^3$. Hence, $\mathbf{a}_\text{MC} \mapsto \ddot{\mathbf{p}}_\text{MC}$ is locally surjective.

\paragraph{Rotational rank}
By \eqref{eq:S_determinant}, $\mathrm{rank}(\mathbf{S}) = 3$, whenever $\cos\phi_\text{G}\cos\theta_\text{G} \neq 0$.

Therefore, $\mathrm{rank}(\mathbf{J}) = 6$ almost everywhere, and the camera pose is \emph{output fully actuated} and \emph{small-time locally controllable}~(STLC) on $SE(3)$. 

\begin{remark}
With a 3-axis gimbal ($\mathbf{S} \equiv \mathbf{I}_3$), the result holds globally. 
\end{remark}

\subsection{Camera-Pose Mimicry}
\label{ssec:mimicry}

Now we show that multicopter and gimbal control inputs exist so that 
\begin{equation}
\mathbf{y}_\text{MC}(t) \equiv \mathbf{y}_\text{AC}(t) \quad \forall t.
\end{equation}

\begin{assumption}[Authority]
\label{ass:authority}
Thrust and torque of the multicopter, as well as gimbal angle/rate limits, are not violated along the trajectories of interest.
\end{assumption}

\paragraph{Position matching}
Equality of camera positions, $\mathbf{p}_\text{M} + \mathbf{R}(\mathbf{q}_\text{M})\mathbf{t}_\text{G} = \mathbf{p}_\text{A} + \mathbf{R}(\mathbf{q}_\text{A})\mathbf{t}_\text{AC}$, is realised by commanding the multicopter position to the lever-arm-compensated target
\begin{equation}
\label{eq:pos_cmd_combined}
\mathbf{p}_\text{M} = \mathbf{p}_\text{A} + \mathbf{R}(\mathbf{q}_\text{A})\mathbf{t}_\text{AC} - \mathbf{R}(\mathbf{q}_\text{M})\mathbf{t}_\text{G}.
\end{equation}
Taking two derivatives gives the desired multicopter's acceleration
\begin{equation}
\mathbf{a}_\text{M} = \ddot{\mathbf{p}}_\text{A} + \frac{d^2 \mathbf{R}(\mathbf{q}_\text{A})}{dt^2} \mathbf{t}_\text{AC} - \frac{d^2 \mathbf{R}(\mathbf{q}_\text{M})}{dt^2} \mathbf{t}_\text{G},
\end{equation}
which the $SE(3)$ inner loop (third line of \eqref{eq:quad_dynamics}) can track so that $\mathbf{p}_\text{M} + \mathbf{R}(\mathbf{q}_\text{M})\mathbf{t}_\text{G} \rightarrow \mathbf{p}_\text{A} + \mathbf{R}(\mathbf{q}_\text{A})\mathbf{t}_\text{AC}$.

\paragraph{Orientation matching}
We require $\mathbf{q}_\text{M} \otimes \mathbf{q}_\text{G} = \mathbf{q}_\text{A}$. Decompose the multicopter attitude into free yaw and tilt:
\begin{equation}
\label{eq:q_split_combined}
\mathbf{q}_\text{M} = \mathbf{q}_z(\psi_\text{M}) \otimes \mathbf{q}_\text{tilt},
\end{equation}
where $\mathbf{q}_\text{tilt}$ sets the thrust direction dictated by $\mathbf{a}_\text{M}$, and $\psi_\text{M}$ is the free yaw about $\vec{\mathbf{z}}_\text{M}$.
Define relative camera quaternion
\begin{equation}
\label{eq:q_rel_combined}
\mathbf{q}_\text{rel} = \bar{\mathbf{q}}_\text{M} \otimes \mathbf{q}_\text{A} = \bar{\mathbf{q}}_\text{tilt} \otimes \mathbf{q}_z(-\psi_\text{M}) \otimes \mathbf{q}_\text{A}.
\end{equation}
To cancel the yaw, choose
\begin{equation}
\label{eq:yaw_sched_combined}
\bar{\psi}_\text{M} = \mathrm{yaw}\left( \bar{\mathbf{q}}_\text{tilt} \otimes \mathbf{q}_\text{A} \right).
\end{equation}
Then, $\mathbf{q}_\text{rel}$ has zero yaw and, therefore, lies in the roll-pitch submanifold. Hence, there exist $(\phi_\text{G}, \theta_\text{G})$ such that
\begin{equation}
\label{eq:qG_choice_combined}
\mathbf{q}_\text{G} = \mathbf{q}_x(\phi_\text{G}) \otimes \mathbf{q}_y(\theta_\text{G}) = \mathbf{q}_\text{rel}.
\end{equation}
This yields $\mathbf{q}_\text{M} \otimes \mathbf{q}_\text{G} = \mathbf{q}_\text{A}$, so camera orientations match exactly.

Under Assumptions~\ref{ass:non_singularity} and \ref{ass:authority}, the multicopter with a \mbox{2-axis} gimbal is output fully actuated and STLC on $SE(3)$ for the camera pose, and control inputs exist which realise $\mathbf{y}_\text{MC}(t) \equiv \mathbf{y}_\text{AC}(t)$ for any feasible fixed-wing camera trajectory. With identical intrinsics, the rendered image streams are indistinguishable to the viewer.

\section{Proposed Method}
\label{sec:method}

Given the fixed-wing camera trajectory $\mathbf{y}_\text{AC}(t)$ in~\eqref{eq:plane_camera_pose}, we want to compute multicopter and gimbal inputs, 
\begin{equation}
\mathbf{u}_\text{MC} = \begin{bmatrix} \mathbf{u}_\text{M} \\ \mathbf{u}_\text{G} \end{bmatrix},
\label{eq:u_MC}
\end{equation}
such that the multicopter camera pose $\mathbf{y}_\text{MC}(t)$ in~\eqref{eq:gimbal_y} satisfies $\mathbf{y}_\text{MC}(t) \equiv \mathbf{y}_\text{AC}(t) \quad \forall t$.

To enable multicopter tracking aggressive manoeuvres, non-linear control techniques must be used. Dynamic feedback linearization~(DFL) could be designed to achieve better convergence and robustness. Through a change of variables, DFL transforms a nonlinear system into an equivalent linear, controllable and observable one.
Given a non-linear system
\begin{equation}
\begin{cases}
\dot{\mathbf{x}} = f(\mathbf{x}) + g(\mathbf{x}) \mathbf{u} \\
\mathbf{y} = h(\mathbf{x}),
\end{cases}
\label{eq:nonlinear_system}
\end{equation}
where $\mathbf{x}$ is the system state, $\mathbf{u}$ is the system input, $\mathbf{y}$ is the system output, $f(\mathbf{x})$, $g(\mathbf{x})$ and $h(\mathbf{x})$ are vector fields in $\mathbb{R}^n$. The task of feedback linearization is to identify a static state feedback control law of the following form:
\begin{equation}
\mathbf{u} = \alpha(\mathbf{x}) + \beta(\mathbf{x}) \mathbf{v},
\label{eq:feedback_control}
\end{equation}
where $\mathbf{v}$ is an new control input, $\alpha(\mathbf{x})$ and $\beta(\mathbf{x})$ are smooth functions defined in a neighbourhood of some point $\mathbf{x}_0 \in  \mathbb{R}^n$ and $\beta(\mathbf{x}_0) \neq 0$, such that the closed-loop system composed of (\ref{eq:nonlinear_system}) and (\ref{eq:feedback_control}) behaves as a linear and completely accessible system.
For system (\ref{eq:nonlinear_system}), let $N_\text{I}$ be the number of inputs, $N_\text{O}$ be the number of outputs, $\mathbf{r} = \begin{bmatrix}r_1 & \cdots & r_{N_\text{O}}\end{bmatrix}$ be the vector relative degree. Then, let's define
\begin{equation}
\Delta(\mathbf{x}) = \begin{bmatrix}
L_{g_1} L^{r_1 - 1}_f h_1(\mathbf{x}) & \cdots & L_{g_{N_\text{I}}} L^{r_1 - 1}_f h_1(\mathbf{x}) \\
\vdots & \ddots & \vdots \\
L_{g_1} L^{r_{N_\text{O}} - 1}_f h_{N_\text{O}}(\mathbf{x}) & \cdots & L_{g_{N_\text{I}}} L^{r_{N_\text{O}} - 1}_f h_{N_\text{O}}(\mathbf{x})
\end{bmatrix}
\label{eq:delta}
\end{equation}
and
\begin{equation}
b(\mathbf{x}) = \begin{bmatrix}L^{r_1}_f h_1(\mathbf{x}) & \cdots & L^{r_{N_\text{O}}}_f h_{N_\text{O}}(\mathbf{x})\end{bmatrix}^T,
\label{eq:b}
\end{equation}
where $L^r_f h$ denotes the $r$-th Lie derivative of the function $h$ with respect to the vector field $f$.
The main result about the input-output decoupling problem is that this problem is solvable if and only if the matrix $\Delta(\mathbf{x})$ is non-singular. In that case, the static state feedback (\ref{eq:feedback_control}) with
\begin{equation}
\begin{cases}
\alpha(\mathbf{x}) = -\left( \Delta(\mathbf{x}) \right)^{-1} b(\mathbf{x}) \\
\beta(\mathbf{x}) = \left( \Delta(\mathbf{x}) \right)^{-1}
\end{cases}
\label{eq:alpha_beta}
\end{equation}
renders the closed-loop system linear and decoupled from an input-output point of view.

To feedback linearize a system, the necessary and sufficient condition for the solvability of the state space exact linearization problem is \cite{Isidori1986CDC}:
\begin{equation}
\sum_{i = 1}^{N_\text{O}}{r_i} = N,
\label{eq:condition}
\end{equation}
where $N$ is the system's rank.

\tikzstyle{int}=[draw, align=center, fill=blue!0, minimum size=2em, text width=1cm]
\tikzstyle{sum}=[draw, fill=blue!0, shape=circle, inner sep=0.3pt]
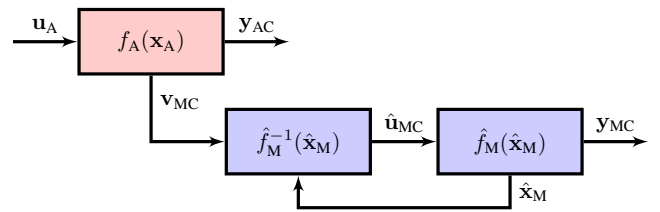
\begin{figure}[!b]
\centering
    \resizebox{0.99\columnwidth}{!}{
    \begin{tikzpicture}[node distance=1cm, auto, >=latex', line width=0.5mm]
        \node[int, text width=2cm, minimum height=1cm, fill=red!20](f_A){$f_\text{A}(\mathbf{x}_\text{A})$};
        |
        \node[int, below right=0.5cm and 0cm of f_A, text width=2cm, minimum height=1cm, fill=blue!20](f_M1){$\hat{f}_\text{M}^{-1}(\hat{\mathbf{x}}_\text{M})$};
        \node[int, right=1cm of f_M1, text width=2cm, minimum height=1cm, fill=blue!20](f_M){$\hat{f}_\text{M}(\hat{\mathbf{x}}_\text{M})$};
   
        \draw[->] ($(f_A.west) + (-1cm, 0)$) -- node{$\mathbf{u}_\text{A}$} (f_A.west);
        \draw[->] (f_A.east) -- node{$\mathbf{y}_\text{AC}$} ++(1cm, 0);
        \draw[->] (f_A.south) |- node[pos=0.2]{$\mathbf{v}_\text{MC}$} (f_M1.west);
        \draw[->] (f_M1.east) -- node{$\hat{\mathbf{u}}_\text{MC}$} (f_M.west);
        \draw[->] (f_M.east) -- node{$\mathbf{y}_\text{MC}$} ++(1cm, 0);
        \draw[->] (f_M.south) -- node{$\hat{\mathbf{x}}_\text{M}$} ($(f_M.south) + (0, -0.5cm)$) -| (f_M1.south);
    \end{tikzpicture}}
\caption{Overview of the proposed framework.}
\label{fig:framework}
\end{figure}

For the system~(\ref{eq:quad_dynamics}), it can be shown that $\mathbf{r} = \begin{bmatrix}2 & 2 & 2 & 1 & 1 & 1\end{bmatrix}$, and $N = 14$ ~\cite{sarabakha2025data}. Since, $\sum_{i = 1}^{6}{r_i} = 9 \neq 14$, the condition~(\ref{eq:condition}) is not satisfied. Therefore, the input-output decoupling problem is not solvable for the system~(\ref{eq:quad_dynamics}) by means of a static state feedback control law. In fact, $\Delta(\mathbf{x}_\text{M})$ is singular for any $\mathbf{x}_\text{M}$ and is not invertible. So, the system (\ref{eq:quad_dynamics}) can not be transformed into an equivalent linear and controllable one by static state feedback. This is due to the fact that $\mathbf{a}_\text{M}$ is affected only by thrust $T$ and not by other control inputs. In order to get $\Delta(\mathbf{x})$ non-singular, the appearance of $T$ should be delayed to higher order derivatives of $\mathbf{a}_\text{M}$. This can be achieved by a dynamic compensator which incorporates an additional set of state variables $\boldsymbol{\chi}$ to achieve higher relative degrees~\cite{Mistler2001ROMAN}.
By setting $T$ equal to the output of an auxiliary dynamic system driven by a new reference input $\hat{T}$. The simplest way in which this result can be achieved is to set this auxiliary dynamic system equal to a double integrator. In this case, $\boldsymbol{\chi} = \begin{bmatrix} \chi_1 & \chi_2 \end{bmatrix}^\top$. So, $T$ is set equal to the output of a double integrator driven by $\ddot{T}$. The other input variables in~\eqref{eq:u_MC} are unchanged.
Now, $T$ is not any-more an input for the system~\eqref{eq:quad_dynamics}, but becomes the internal state for the new dynamical system
\begin{equation}
\hat{f}_\text{M}:
\begin{cases}
\dot{\mathbf{p}}_\text{M} = \mathbf{R}\left(\mathbf{q}_\text{M}\right) \mathbf{v}_\text{M} \\
\dot{\mathbf{q}}_\text{M} = \frac{1}{2} \mathbf{G}\left(\mathbf{q}_\text{M}\right) \boldsymbol{\omega}_\text{M} \\
\dot{\mathbf{v}}_\text{M} = -\boldsymbol{\omega}_\text{M} \times \mathbf{v}_\text{M} + \frac{1}{m_\text{M}} \begin{bmatrix} 0 & 0 & \chi \end{bmatrix}^\top \\
\dot{\boldsymbol{\omega}}_\text{M} = -\left(\mathbf{J}_\text{M}\right)^{-1} \boldsymbol{\omega}_\text{M} \times (\mathbf{J}_\text{M} \boldsymbol{\omega}_\text{M}) + \left(\mathbf{J}_\text{M}\right)^{-1} \boldsymbol{\tau}_\text{M} \\
\dot{\mathbf{q}}_\text{MC} = \frac{1}{2} \mathbf{G}\left(\mathbf{q}_\text{MC}\right) \mathbf{u}_\text{G} \\
\ddot{\chi} = \ddot{T}
\end{cases}.
\label{eq:new_quad_dynamics}
\end{equation} 
The new control inputs to the system~\eqref{eq:new_quad_dynamics} are 
\begin{equation}
\hat{\mathbf{u}}_\text{MC} = \begin{bmatrix}\ddot{T} & \tau_{\phi} & \tau_{\theta} & \tau_{\psi} & \dot\theta_\text{G} & \dot\psi_\text{G} \end{bmatrix}^\top.
\end{equation}
The input-output decoupling problem is solvable for the system (\ref{eq:quad_dynamics}) by means of a dynamic feedback control law, if it is solvable via static feedback for the extended system (\ref{eq:new_quad_dynamics}). Since the extended system has $\hat{\mathbf{r}} = \begin{bmatrix} 4 & 4 & 4 & 2 & 1 & 1 \end{bmatrix}$ and $\hat{N} = 16$, the condition~\eqref{eq:condition} is satisfied. Thus, the input-output decoupling problem is solvable for the system (\ref{eq:new_quad_dynamics}) by means of a dynamic feedback control law as in~\eqref{eq:feedback_control}:
\begin{equation}
\hat{\mathbf{u}}_\text{MC} = \alpha(\hat{\mathbf{x}}_\text{MC}) + \beta(\hat{\mathbf{x}}_\text{MC}) \mathbf{v}_\text{MC},
\label{eq:feedback_control_extended}
\end{equation}
where $\alpha(\hat{\mathbf{x}}_\text{MC})$ and $\beta(\hat{\mathbf{x}}_\text{MC})$ are computed using~\eqref{eq:alpha_beta}, and
\begin{equation}
\mathbf{v}_\text{MC} = \begin{bmatrix} \mathbf{p}_\text{AC}^\top & \dot{\mathbf{p}}_\text{AC}^\top & \ddot{\mathbf{p}}_\text{AC}^\top & \dddot{\mathbf{p}}_\text{AC}^\top & \ddddot{\mathbf{p}}_\text{AC}^\top & \mathbf{q}_\text{A}^\top & \dot{\mathbf{q}}_\text{A}^\top & \ddot{\mathbf{q}}_\text{A}^\top \end{bmatrix}.
\end{equation}
An overview of the proposed framework is shown in Fig. \ref{fig:framework}.




\begin{figure*}[b]
    \centering
    \includegraphics[width=\textwidth]{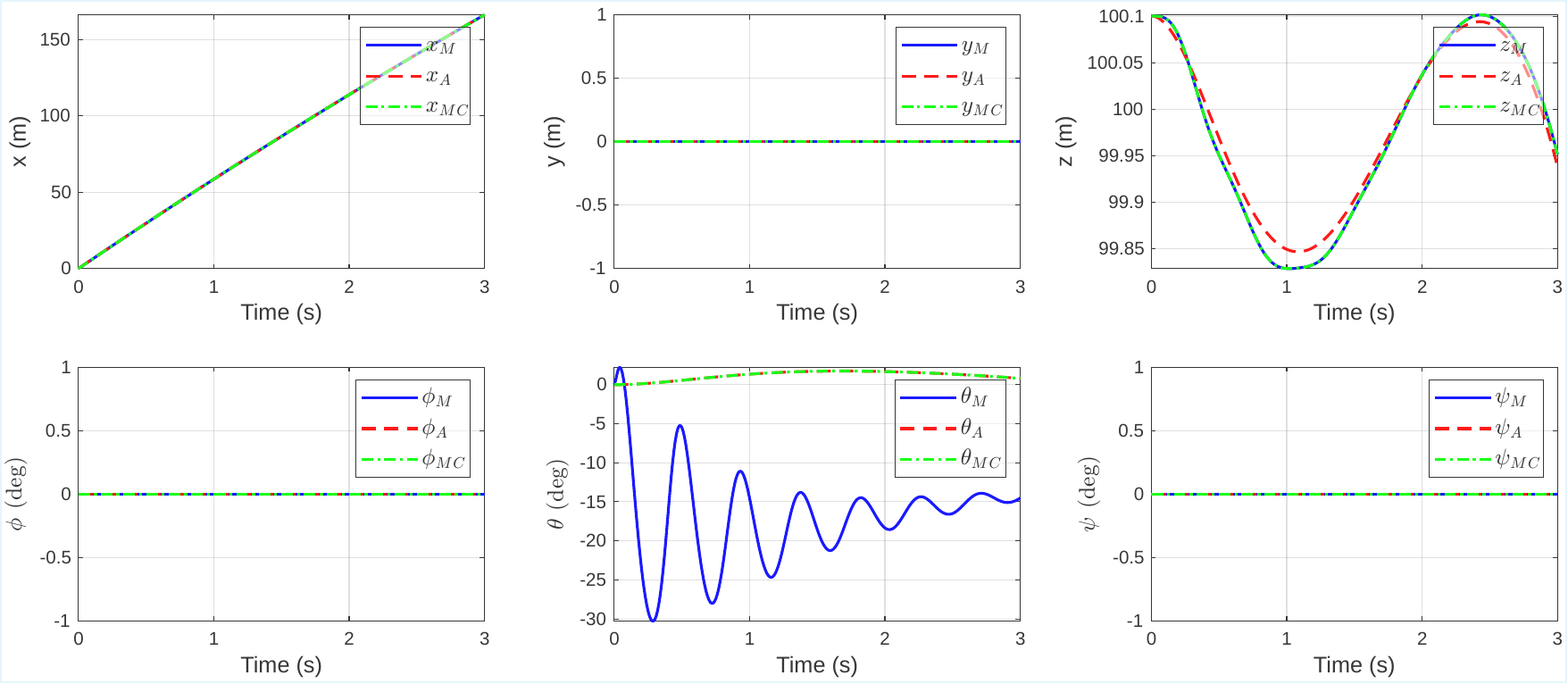}
    \caption{Position and orientation tracking during level-flight.}
    \label{fig:level_tracking}
\end{figure*}

\section{Validation Results}
\label{sec:results}

To validate the proposed DFL controller for camera-pose mimicry, we simulate a multicopter equipped with a 2-axis gimbal tracking the flight of a Red Bull Edge 540~\cite{redbull} aerobatic aircraft. The Edge 540 was selected for its aggressive maneuvering capabilities, providing a challenging test case for the controller. The multicopter chosen is the Freefly Alta-X platform \cite{FreeflyAltaX_spec}, featuring a high thrust-to-weight ratio necessary for tracking high-speed trajectories while maintaining control authority as required by Assumption~\ref{ass:authority}. The parameters are provided in Table~\ref{tab:vehicle_params}.

\begin{table}[!b]
\centering
\caption{Vehicle Parameters}
\label{tab:vehicle_params} 
\begin{tabular}{|l|l|}
\hline
\multicolumn{2}{|c|}{\textbf{Fixed-Wing (Redbull Edge 540)}} \\
\hline
Mass & 530 kg \\
Moments of inertia & (2419, 2133, 4552) kg·m² \\
Wingspan & 7.40 m \\
Wing Area & 9.10 m² \\
Mean Chord & 1.23 m \\
Max Speed & 118 m/s \\
Lift & 0.20, 4.0, 0.0, -0.36 \\ 
Drag & 0.03, 0.04, 0.0, 0.0, 0.0 \\ 
Pitch Moment & -0.01, -0.5, -3.6, -0.5 \\ 
Side Force & -0.98, 0.0, 0.0, 0.0, -0.17 \\
Rolling & -0.12, -0.26, 0.14, 0.08, 0.105 \\
Yaw Moment & 0.25, 0.022, -0.35, 0.06, -0.032 \\
\hline
\multicolumn{2}{|c|}{\textbf{Multicopter (Freefly Alta-X)}} \\
\hline
Mass & 10.6 kg \\
Moments of inertia & (0.48, 0.48, 0.41) kg·m² \\
\hline
\end{tabular}%
\end{table}

Two distinct flight scenarios are examined: (1) a level flight trajectory, and (2) an aerobatic loop manoeuvre testing the controller's ability to handle aggressive dynamics. For both scenarios, we evaluate the system's ability to achieve the mimicry condition $\mathbf{y}_\text{MC}(t) \equiv \mathbf{y}_\text{AC}(t)$ established in Section~\ref{sec:proof}.

\subsection{Trajectory 1: Level Flight}

\begin{figure}[!b]
    \centering
    \includegraphics[width=0.33\textwidth]{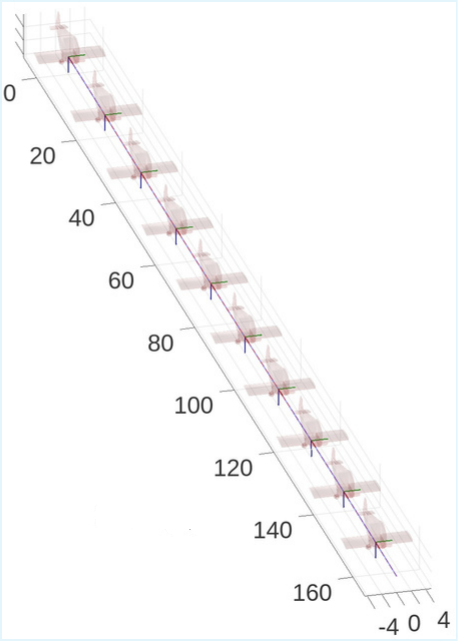}
    \caption{3D trajectories of the fixed-wing aircraft (red), multicopter position (blue), and gimbal camera position (3-D axis) for level flight.}
    \label{fig:3d_trajectory_level}
\end{figure}

\begin{figure*}[b]
    \centering
    \includegraphics[width=\textwidth]{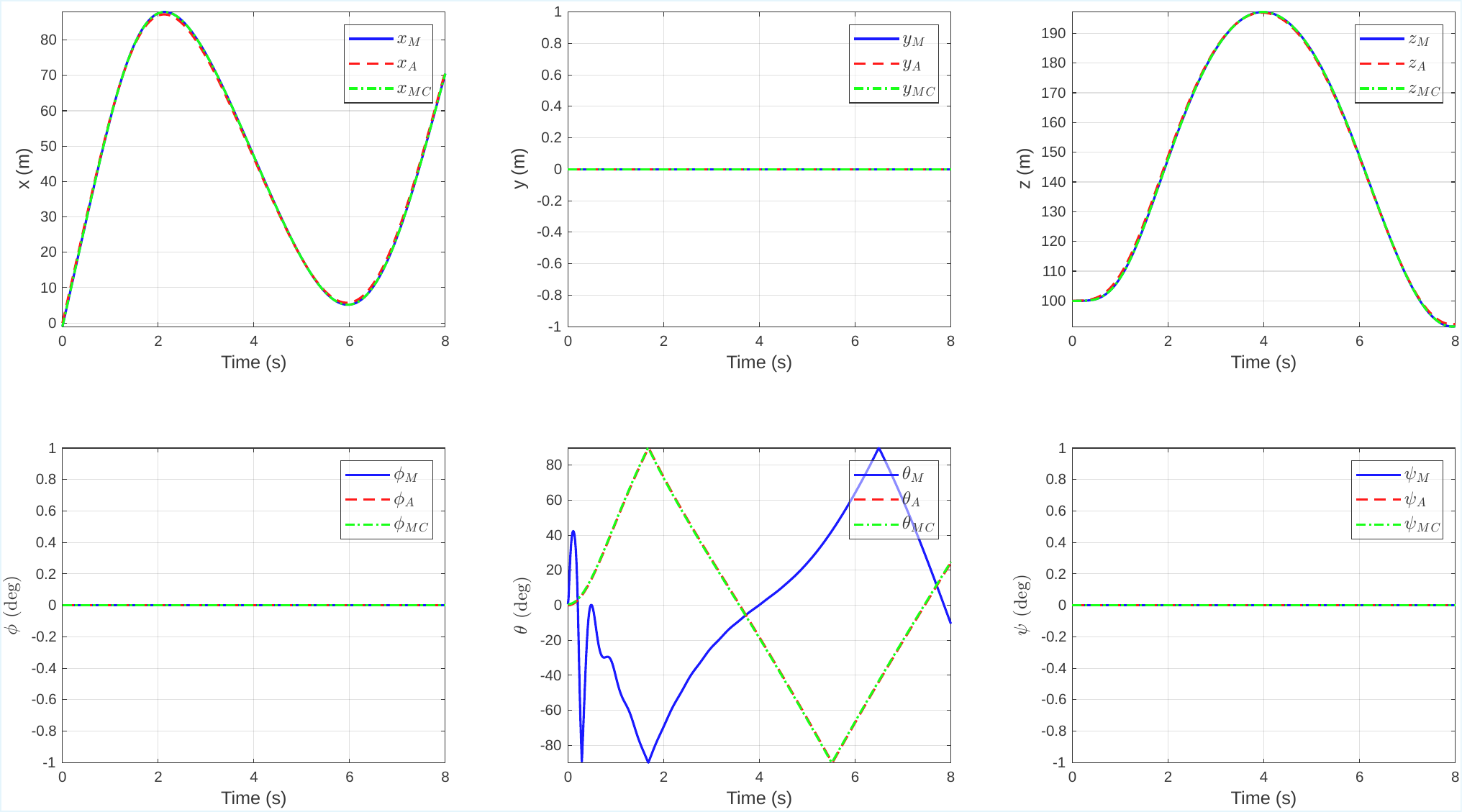}
    \caption{Position and orientation tracking during the loop manoeuvre. Despite the aggressive dynamics, the gimbal camera position and orientation (green) accurately follow the fixed-wing reference (red), while the multicopter (blue) exhibits a large pitch necessary to maintain the required thrust vectoring. }
    \label{fig:loop_tracking}
\end{figure*}

The level flight trajectory serves as a baseline validation of the controller's tracking performance under quasi-steady conditions. The fixed-wing aircraft maintains level flight with a constant elevator deflection of $\delta_e = -0.05$ rad to counteract gravity. 

Fig.~\ref{fig:level_tracking} provides detailed tracking performance across all six degrees of freedom. The position plots (top panels) confirm that the gimbal camera position tracks the fixed-wing reference with negligible error, validating the translational component of the mimicry condition. The orientation plots (bottom panels) reveal the key insight of the proposed approach: while the multicopter body (blue) pitches to angles reaching 30° to generate the forward thrust required to match the fixed-wing's speed, the gimbal (green) actively compensates by counter-rotating, ensuring that the camera orientation precisely follows the fixed-wing's attitude (red). This behavior directly demonstrates the pose matching strategy presented in Section~\ref{sec:method}, where the control dynamic feedback control $\hat{\mathbf{u}}_\text{MC}$ is computed via Eq.~\eqref{eq:feedback_control_extended} to satisfy $\mathbf{q}_\text{M} \otimes \mathbf{q}_\text{G} = \mathbf{q}_\text{A}$.

The rapid gimbal response and absence of steady-state error confirm that the system operates well within the non-singular region defined by Assumption~\ref{ass:non_singularity} ($\cos\phi_\text{G}\cos\theta_\text{G} \neq 0$), maintaining full actuation authority throughout the trajectory.

\subsection{Trajectory 2: Aerobatic Loop}

To stress-test the DFL controller under highly dynamic conditions, we command the fixed-wing aircraft to perform a vertical loop with a constant elevator deflection of $\delta_e = -0.25$ rad. This maneuver generates rapid variations in position, velocity, and orientation, with the aircraft experiencing significant changes in airspeed and load factor throughout the loop.

\begin{figure}[!b]
    \centering
    \includegraphics[width=0.4\textwidth]{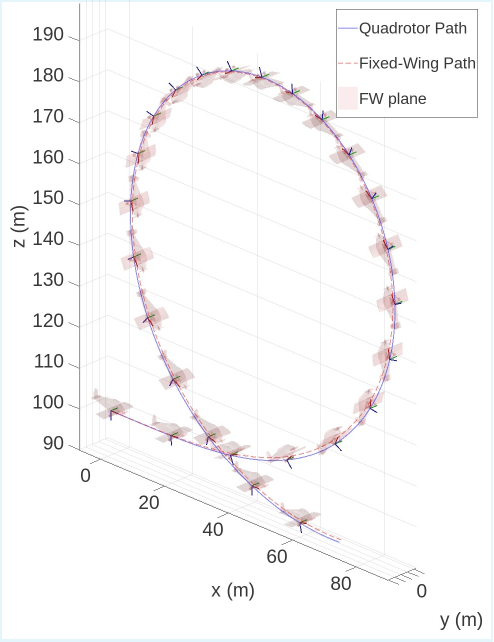}
    \caption{3D trajectories of the fixed-wing aircraft (red), multicopter position (blue), and gimbal camera position (3-D axis) for loop manoeuvres.}
    \label{fig:3d_trajectory_loop}
\end{figure}

As evident in Fig.~\ref{fig:3d_trajectory_loop}, the multicopter-gimbal system successfully completes the loop maneuver with the camera trajectory (green) closely following the fixed-wing path (red). The three-dimensional visualization demonstrates the controller's ability to handle rapid directional changes and varying gravitational loads throughout the vertical plane maneuver.

Fig.~\ref{fig:loop_tracking} shows the full complexity of the tracking task. During the loop, the multicopter body orientation (blue) undergoes extreme pitch variations, exceeding 60° at the top of the loop, as the controller commands the thrust vector to track the aircraft's curved trajectory while counteracting changing gravitational forces. Simultaneously, the gimbal executes rapid compensatory motions to decouple the camera orientation from these aggressive multicopter body rotations. The result is that the gimbal camera orientation (green) remains tightly coupled to the fixed-wing reference (red), with tracking errors remaining bounded despite the challenging dynamics.

The position tracking (top panels of Fig.~\ref{fig:loop_tracking}) shows brief transients during the highest-acceleration phases of the loop, but the controller maintains stability and convergence throughout. This performance validates the feedback linearization approach developed in Section~\ref{sec:method}, where the dynamic compensator (Eq.~\eqref{eq:new_quad_dynamics}) extends the relative degree to enable exact input-output decoupling. The linearized dynamics allow the controller to compute the required thrust derivatives $\ddot{T}$ and torques $\boldsymbol{\tau}_\text{M}$ that drive the tracking errors to zero.

\subsection{Discussion}
The simulation results validate the theoretical framework presented in Sections~\ref{sec:proof} and \ref{sec:method}. The multicopter-gimbal system successfully achieves camera-pose mimicry $\mathbf{y}_\text{MC}(t) \equiv \mathbf{y}_\text{AC}(t)$ for both steady and aggressive trajectories, confirming output full actuation on $SE(3)$, effective attitude decoupling, and the efficacy of the DFL controller with dynamic compensation in achieving precise tracking of the nonlinear, coupled multicopter-gimbal dynamics. The high-amplitude pitch oscillations observed in the multicopter body orientation during both trajectories, especially pronounced in the loop maneuver, highlight the inherent challenge of fixed-wing mimicry, where the multicopter must continuously reorient its thrust vector to follow the aircraft’s velocity profile. The gimbal’s ability to maintain smooth camera orientation despite these aggressive motions demonstrates the practical value of the proposed architecture.

\section{Conclusions and Future Work}
\label{sec:conclusions}

This work has designed and validated a control framework for physical-to-physical emulation, demonstrating that a multicopter equipped with a two-axis gimbal can dynamically replicate the camera viewpoint and flight dynamics of a fixed-wing aircraft. Our approach, based on DFL, provides a practical method for achieving high-fidelity mimicry of a primary system's key observables. This capability enables algorithm validation and operator training in real-world conditions at a negligible cost and with significantly reduced operational risk compared to deploying the high-value fixed-wing planes. Furthermore, the principles developed can be generalized to other classes of dissimilar dynamic systems, such as ground or underwater vehicles, opening new paradigms for cross-system validation.


\bibliographystyle{IEEEtran}
\bibliography{References}

@IEEEtranBSTCTL{IEEEexample:BSTcontrol,
CTLdash_repeated_names = "no"
}

@INPROCEEDINGS{Isidori1986CDC,
  author={Isidori, A. and Moog, C. H. and Luca, A. De},
  booktitle={1986 25th IEEE Conference on Decision and Control}, 
  title={{A sufficient condition for full linearization via dynamic state feedback}}, 
  year={1986},
  volume={},
  number={},
  pages={203-208},
  doi={10.1109/CDC.1986.267208}}

@INPROCEEDINGS{Mistler2001ROMAN,
  author={Mistler, V. and Benallegue, A. and M'Sirdi, N.K.},
  booktitle={Proceedings 10th IEEE International Workshop on Robot and Human Interactive Communication. ROMAN 2001 (Cat. No.01TH8591)}, 
  title={Exact linearization and noninteracting control of a 4 rotors helicopter via dynamic feedback}, 
  year={2001},
  volume={},
  number={},
  pages={586-593},
  doi={10.1109/ROMAN.2001.981968}}

@article{yuan2025design,
  author    = {X. Yuan and J. Xu and S. Li},
  title     = {Design and Simulation Verification of Model Predictive Attitude Control Based on Feedback Linearization for Quadrotor UAV},
  journal   = {Applied Sciences},
  volume    = {15},
  number    = {9},
  pages     = {5218},
  year      = {2025}
}

@article{wu2019modeling,
  author    = {H. Wu and Z. Wang and Z. Zhou and R. Wang},
  title     = {Modeling and Simulation for Multi-Rotor Fixed-Wing UAV Based on Multibody Dynamics},
  journal   = {Journal of Northwestern Polytechnical University},
  volume    = {37},
  number    = {5},
  pages     = {928--934},
  year      = {2019}
}

@misc{wang2023similarity,
  title   = {Similarity Between Two Dynamical Systems},
  author  = {Wang, X. and Li, Y. and Han, Y.},
  year    = {2023},
  eprint  = {2310.03383},
  archivePrefix = {arXiv},
  primaryClass = {math.DS},
  url     = {https://arxiv.org/abs/2310.03383}
}

@inproceedings{amer2023unav,
  title={Unav-sim: A visually realistic underwater robotics simulator and synthetic data-generation framework},
  author={Amer, Abdelhakim and {\'A}lvarez-Tu{\~n}{\'o}n, Olaya and U{\u{g}}urlu, Halil {\.I}brahim and Sejersen, Jonas Le Fevre and Brodskiy, Yury and Kayacan, Erdal},
  booktitle={2023 21st International Conference on Advanced Robotics (ICAR)},
  pages={570--576},
  year={2023}
}

@inproceedings{amer2023visual,
  title={Visual Tracking Nonlinear Model Predictive Control Method for Autonomous Wind Turbine Inspection},
  author={Amer, Abdelhakim and Mehndiratta, Mohit and le Fevre Sejersen, Jonas and Pham, Huy Xuan and Kayacan, Erdal},
  booktitle={2023 21st International Conference on Advanced Robotics (ICAR)},
  pages={431--438},
  year={2023},
  organization={IEEE}
}

@article{amer2025,
  title={Empowering Autonomous Underwater Vehicles Using Learning-Based Model Predictive Control With Dynamic Forgetting Gaussian Processes},
  author={Amer, Abdelhakim and Mehndiratta, Mohit and Brodskiy, Yury and Kayacan, Erdal},
  journal={IEEE Transactions on Control Systems Technology},
  year={2025},
  publisher={IEEE}
}

@misc{redbull,
  author = {David C. Eyre},
  title = {Zivko Edge 540 — Aircraft model},
  year = {2010},
  note = {“Encyclopedia of Aircraft” web article, wingspan = 7.62 m, wing area = 9.10 m$^2$, rate of roll = 420°/s, etc.},
  url = {https://aeropedia.com.au/content/zivko-edge-540/}
}

@misc{FreeflyAltaX_spec,
  author       = {Freefly Systems},
  title        = {Alta X — Technical Specifications},
  year         = {2025},
  note         = {Unfolded diameter 1.415 m (frame), max gross take-off weight 34.9 kg, max payload 15.9 kg.},
  url          = {https://freeflysystems.com/alta-x/specs}
}

@inproceedings{sarabakha2025data,
  title={Data-Driven Identification of Observed Relative Degrees for Nonlinear Systems},
  author={Sarabakha, Andriy},
  booktitle={2025 European Control Conference (ECC)},
  pages={813--818},
  year={2025},
  organization={IEEE}
}

@article{smith2022emulation,
  title   = {A Review of Dynamic Emulation for Validation of Autonomous Systems},
  author  = {Smith, J. and Anderson, B.},
  journal = {Journal of Field Robotics},
  year    = {2022},
  volume  = {39},
  number  = {2},
  pages   = {145--168}
}

@inproceedings{gao2020realitygap,
  title     = {The Reality Gap: A Survey of Sim-to-Real Transfer in Robotics},
  author    = {Gao, Y. and Lin, J. and Wu, F.},
  booktitle = {2020 IEEE International Conference on Robotics and Automation (ICRA)},
  year      = {2020},
  pages     = {5678--5684},
  doi       = {10.1109/ICRA40945.2020.9197115}
}

@article{pshuniak2018relative,
  title   = {Relative Degree and {DFL}-based Control of a {UAV} with a Suspended Load},
  author  = {Pshuniak, A. and Korpela, C. and Mohyla, V.},
  journal = {Journal of Intelligent \& Robotic Systems},
  year    = {2When018},
  volume  = {90},
  number  = {1},
  pages   = {201--219}
}

@inproceedings{shah2017airsim,
  title={Airsim: High-fidelity visual and physical simulation for autonomous vehicles},
  author={Shah, Shital and Dey, Debadeepta and Lovett, Chris and Kapoor, Ashish},
  booktitle={Field and service robotics: Results of the 11th international conference},
  pages={621--635},
  year={2017},
  organization={Springer}
}

@inproceedings{edvani1997simona,
  title={Simona-a reconfigurable and versatile research facility},
  author={Edvani, Sunjoo and va, Etienne and Bettendorf, Sander and Edvani, Sunjoo and va, Etienne and Bettendorf, Sander},
  booktitle={Modeling and Simulation Technologies Conference},
  pages={3809},
  year={1997}
}

@inproceedings{stroosma2003using,
  title={Using the SIMONA research simulator for human-machine interaction research},
  author={Stroosma, Olaf and Van Paassen, MM and Mulder, Max},
  booktitle={AIAA modeling and simulation technologies conference and exhibit},
  pages={5525},
  year={2003}
}

@article{aljalbout2025reality,
  title={The Reality Gap in Robotics: Challenges, Solutions, and Best Practices},
  author={Aljalbout, Elie and Xing, Jiaxu and Romero, Angel and Akinola, Iretiayo and Garrett, Caelan Reed and Heiden, Eric and Gupta, Abhishek and Hermans, Tucker and Narang, Yashraj and Fox, Dieter and others},
  journal={arXiv preprint arXiv:2510.20808},
  year={2025}
}

\end{document}